\documentstyle[preprint,aps]{revtex}
\begin{document}
\tightenlines
\title{Explosions of water clusters in intense laser fields}
\author{V. Kumarappan\footnote{Now at: Institute for Physical Sciences and Technology, University of Maryland, Maryland, USA}, M. Krishnamurthy, and D. Mathur} 
\address{Tata Institute of Fundamental Research, 1 Homi Bhabha 
Road, Mumbai 400 005, India.}
\date{\today}
\maketitle
\begin{abstract}
Energetic, highly-charged oxygen ions, $O^{q+}$ ($q\leq 6$), are copiously produced upon laser field-induced disassembly of highly-charged water clusters, $(H_2O)_n$ and $(D_2O)_n$, $n\sim$ 60, that are formed by seeding high-pressure helium or argon with water vapor. $Ar_n$ clusters (n$\sim$40000) formed under similar experimental conditions are found undergo disassembly in the Coulomb explosion regime, with the energies of $Ar^{q+}$ ions showing a $q^2$ dependence. Water clusters, which are argued to be considerably smaller in size,  should also disassemble in the same regime, but the energies of fragment O$^{q+}$ ions are found to depend linearly on $q$ which, according to prevailing wisdom, ought to be a signature of hydrodynamic expansion that is expected of much larger clusters. The implication of these observations on our understanding of the two cluster explosion regimes, Coulomb explosion and hydrodynamic expansion, is discussed. Our results indicate that charge state dependences of ion energy do not constitute an unambiguous experimental signature of cluster explosion regime. 
\end{abstract}
\pacs{36.40.Wa, 52.50.Jm, 52.38.Ph, 36.40.Gk}

\section{Introduction} 

A number of scientific and technological reasons have been responsible for a recent resurgence of interest in studies of the interaction of very intense optical fields with large, gas-phase clusters of atoms. It is now established that the mesoscopic nature of cluster dynamics in intense, ultrashort laser fields results in features that are unique to clusters, observed neither in the interaction of low-density matter (atoms and molecules) with fields of similar magnitude and duration, nor in that of high-density matter (solids). For instance, the extraordinarily high degree of efficiency with which clusters absorb the incident laser energy finds no analog in other gaseous matter. Energy deposition rates of up to 240 mW per atom have been measured \cite{boyer} in clusters comprising several hundreds of thousands of atoms. Primarily, it is the electrons in the cluster that directly absorb such energy, but the energy is rapidly redistributed in the form of intense incoherent radiation and very energetic ions. The level of ionization and the mean energy of the electrons are both significantly higher than expected from laser-field-induced ionization of isolated, gas-phase atoms and molecules. The hot cluster expands and breaks up in a few picoseconds; this rapid disassembly results in the emission of ions with kinetic energies that can be as high as 1 MeV \cite{ditmire1}.  These are up to five orders of magnitude larger than the energies that are obtained upon Coulomb explosion of multiply-charged molecules \cite{physrep}.
 
The motivation for research in this field has been two-fold, scientific and technological. From the scientific point of view, we have noted above that clusters in intense fields are intermediate to atoms and small molecules on the one hand, and solids on the other. The transition from single-atom behavior to bulk properties is complex enough to merit investigation for its own sake. For clusters that are not too large, it is also possible for a substantial fraction of electrons to leave the cluster early in its break-up, leaving behind a hot, non-equilibrium and non-neutral plasma, the evolution of which can be strongly influenced by radiation. Such phenomena are of interest in astrophysics, and cluster plasmas offer one of the few avenues available for studying them in the laboratory. From a technological point of view, laser-cluster interactions offer tantalizing possibilities of developing table-top charged particle accelerators. Clusters have also emerged as one of the promising sources of extreme ultraviolet (EUV) radiation for lithography. EUV lithography is the favored approach to the next generation of chip manufacturing technology, where features on the integrated circuit will be less than 100 nm wide. Much research effort is being devoted to improving the yield of EUV radiation so that sufficient emission around 13.5 nm, the wavelength of choice for this application, can be generated \cite{euv}. Apart from incoherent radiation, clusters have shown promise as a source for high-harmonic generation, wherein extremely non-linear processes result in the emission of coherent radiation at large multiples of the laser frequency \cite{donnelly}.  The interaction of laser pulses from tabletop lasers with deuterium and deuterium-rich clusters has also made it possible to study hot-plasma fusion in small facilities, which was hitherto restricted to very large national facilities \cite{ditmire2}.  The fusion process also holds promise as a source of mono-energetic and short-pulse neutrons for medical and material science studies \cite{medical}.
 
Although various theoretical models have been proposed to explain the evolution of the cluster under intense field irradiation, none can adequately explain all the experimentally observed features that presently drive research in this area. It is known that most atoms in a cluster get tunnel ionized at the leading edge of the incident laser pulse. As the ionized electrons leave the cluster, what gets left behind is a positively charged core that gives rise to an increasing potential barrier to further removal of electrons. The question of whether the barrier is sufficient to retain a large fraction of electrons or not is a major point of contention between the two major models of laser-cluster interaction. In the hydrodynamic expansion model, it is assumed that retention of most of the electrons by the cluster results in a spherically symmetric plasma of uniform density. The retained electrons absorb energy from the laser by collisional inverse bremsstrahlung. The hot electron plasma expands due to hydrodynamic pressure, and transfers energy to the ions. The expansion velocity of the plasma is determined by the plasma sound speed, and since the ion and electron charge clouds expand at the same speed, the ions are expected to be significantly hotter than the electrons. In the Coulomb explosion-ionization ignition model of the dynamics, on the other hand, electrons leave the cluster rapidly after tunnel ionization. As a result, there is a build-up of charge on the cluster that gives rise to a radial field that can become large enough to drive further ionization at the surface of the sphere. The removal of these electrons increases the radial field further and ``ignites" ionization. The cluster then explodes due to the Coulombic repulsion between the positively charged ions. 

Although most researchers in the field still regard the details of a cluster explosion as a matter of some debate, at the most basic level, prevailing wisdom tends to indicates that the hydrodynamic expansion model is expected to hold for clusters that are large enough in size to retain a substantial fraction of the electrons, while the ionization ignition-Coulomb explosion model requires the prompt removal of most of the electrons, which would be the case for small clusters. This may be rationalized by considering the cluster diameter $a$ in terms of parameters that are of utility in characterizing plasmas, such as the electron skin depth, $\delta_e$ $(\delta_e=\omega_p/c$, where $\omega_p$ is the plasma frequency), and electron excursion length due to the action of the ponderomotive potential, $\zeta_e$ \cite{medical}. When the electron skin depth is greater than the cluster size ($\delta_e>a$), the laser field can penetrate the interior regions of the cluster. If, under such circumstances, the laser intensity is also high enough to fulfill the relation $\zeta_e>>a$, most ionized electrons would leave the vicinity of the cluster within one optical cycle, thereby setting up a large electrostatic field that, in turn, would lead to further electron heating. This is the ionization ignition-Coulomb explosion scenario. On the other hand, when the cluster size fulfills the conditions $a>>\zeta_e$ and $a>>\delta_e$, the hydrodynamic expansion situation is expected to prevail.

We present in the following results of experiments that we have conducted on water clusters using a two-dimensional time-of-flight technique that enables us to probe the energy and charge state spectrum of ions produced upon cluster explosion. Analysis of the measured dependence of ion energy on ion charge state shows a linear relationship that might have been indicative of the explosion dynamics proceeding in the hydrodynamic regime. However, the water clusters that are produced in our experiments are certainly not large enough to warrant invocation of the hydrodynamic regime. Instead, our results appear to offer experimental evidence that, contrary to expectations on the basis of prevailing wisdom on the dynamics of cluster disassembly in intense laser fields, charge state dependences of ion energy do not constitute an experimental signature of explosion regime. 

\section{Experimental method}

The fact that irradiation of clusters by intense laser radiation invariably results in formation of energetic, highly-charged ions precludes the use of conventional mass spectrometry. Sector instruments would need to employ unrealistically high values of magnetic field in order to carry out momentum discrimination of the type of ions that are produced in such interactions. Radio frequency (rf) methods would also require unacceptably high amplitudes of rf voltages. We show later that conventional time-of-flight techniques are also of limited utility. 

We present in the following results of experiments in which a new two-dimensional time-of-flight (2D-TOF) method is utilized to probe the dynamics of water clusters in femtosecond-duration laser fields of intensity in the range 10$^{14}$-10$^{16}$ W cm$^{-2}$. Conventional Wiley-McLaren time-of-flight (WM-TOF) methods cannot reliably be used due to the extremely large range of ion energies that are produced in the laser-cluster interactions. The WM-TOF geometry ensures space and time focusing such that the resulting mass resolution that is achieved is high; in order to achieve this, it is necessary that the intrinsic spread in the initial velocity of ions of given m/q (mass-to-charge ratio) is small. In the cluster explosion experiments of the type we report here, the initial ion velocities span a range that can cover 5-6 orders of magnitude. Though spatial focusing is intrinsically achieved as the ions are produced in a small and well defined focal volume, the spread in ion velocities and charge states is large enough to make regular WM-TOF experiments impossible. In such  experiments, therefore, one might resort to simple arrival time measurements to determine the velocity of the ions. Such measurements, do not resolve charge of the ions and, in the case of hetero-nuclear clusters, the additional problem of distinguishing different atomic species becomes very difficult. To overcome this experimental problem, arrival time spectra have been measured in conjunction with a magnetic deflection TOF spectrometer in one recent measurement involving argon and xenon clusters \cite{normand}.  As noted by these authors, direct measurements of ion energies are difficult, particularly for the more energetic components in the total spectrum. To analyze the experimental TOF data in order to deduce ion energies, the authors had to compare measured data with results of numerical simulations. Though the charge states of the atomic ions were, indeed, deciphered in these experiments, it is not very clear if the technique can be applied to hetero-nuclear clusters except, perhaps, for clusters like HI$_n$, where the mass difference between the two constituents is very large and protons may be separated from I$^{q+}$ ions by distinguishing the pulse height of the signal produced by the microchannel plate (MCP) detectors. However, this technique is obviously limited since MCP's are intrinsically not very good as pulse height discriminators; with a wide range of ion velocities the experimental limitations become even larger.

The approach we have adopted in our 2D-TOF method is different, relatively simple to implement, and it demonstrates that these experimental problems can be better addressed. We take recourse to one simplification afforded by the very energetic nature of the laser-cluster interaction that we are interested in probing. It is established, and will be shown in the following, that the laser deposits several keV, or more, energy per ion in the cluster. On the other hand, the cluster itself is bound by very weak van der Waal's forces. The binding energy of these forces is, at most, a few millielectron volts for rare gas clusters. Hence, it is a reasonable expectation that the ionic products of the laser-cluster interaction will almost exclusively be atomic ions that exhibit no clustering. The masses of all these (atomic) ions (in the homonuclear clusters) are therefore the same, and velocity measurements directly yield the energy spectrum. This spectrum of the arrival times of atomic ions is one-dimensional TOF spectrometry. On the other hand, the velocity (energy) information that is readily forthcoming from such measurements alone does not reveal anything about the charge state distribution within the plasma that is produced in the course of the laser-cluster interaction. A two-dimensional experimental approach is required to obtain this information, as is described in the following.

The experimental set-up used in our measurements is schematically depicted in Figure 1. A pulsed expansion valve with a supersonic nozzle was used for producing clusters of large size. The nozzle was of 500 $\mu$m diameter, and could be backed up with gas pressures up to 14 bar. The solenoid-activated valve was capable of being operated at up to 50 Hz repetition rate, and the duration of the pulse was variable over 0.1-1 ms. By triggering the pulsed valve and the laser externally, the cluster beam and the laser beam was made to temporally coincide, with a relative jitter of much less than 1 $\mu$s. The large gas load produced by the valve was handled by a 3000 $\ell$ s$^{-1}$ diffusion pump backed by two 1200 $\ell$ m$^{-1}$ rotary pumps. In our experiments with $Ar_n$ ($n>$5000) clusters, the pulsed valve was operated at repetition rates of less than 10 Hz in order to keep the background pressure to workably low values.  As the gas expanded from the nozzle into our vacuum chamber, adiabatic cooling facilitated the formation of large clusters. Confirmation of the formation of clusters and an estimation of their size was carried out by means of Rayleigh scattering, which depends strongly on the mean size of the scatterer. Details of this aspect of the methodology adopted by us have been presented elsewhere \cite{xray} but the salient features of the methodology are summarized in the following. 

Although the experiments on cluster disassembly dynamics that are reported here were conducted using intense 806 nm light, we used 355 nm radiation produced from the third harmonic of a picosecond Nd:YAG laser in order to carry out the light scattering measurements for estimation of the sizes of rare gas clusters. The use of short-wavelength light is, of course, appropriate since the scattering signal scales as $\lambda^{-4}$. The ultraviolet (UV) light was crossed $\sim$2-3 mm from the supersonic jet nozzle shown in Fig. 1. In light scattering experiments, since the scattered light at the same wavelength is probed, and its intensity is very low, stray light from the incident laser tends to overwhelm the signal. We took care that stray light was kept to a minimum by using Brewster windows at laser entry and exit ports, by blackening the walls of the chamber, and by a system of optical baffles. Scattered light was imaged onto a photomultiplier by an f/3.3 lens. The photomultiplier signal was read by a digital storage oscilloscope that was coupled, via a fast bus, to a computerized data acquisition system.  Rayleigh scattering signals of easily measurable intensity appeared for Ar stagnation pressures greater than 2 bar. The variation of the intensity of the scattered light ($S$) as a function of stagnation pressure ($P$) was measured. The scattered signal scales as $S \propto P^3$. We employed the Hagena scaling law \cite{hagena} to estimate the mean cluster size, $N$, using the empirical parameter, 
$\Gamma^*$,
\begin {equation}
 N \approx A (\Gamma^*)^{1.95}.
\end{equation}
\begin {equation}
\Gamma^* = k {\textrm{P}[\textrm{mbar}] \over (T_o[\textrm{K}])^{2.29}} 
{\Big(}{ \phi[\mu m] \over \textrm{tan}~ \alpha}{\Big )}^{0.85},
\end {equation}
where $k$ = 1700 (for Ar), $T_o$=298 K, $\alpha$ is the half opening angle of the jet, $\phi$ is the nozzle diameter and 
\textit{A} is an empirical constant determined from Ref.\cite{farges}. 
Using the facts that (i) Rayleigh scattering efficiency scales as $r^6$, where $r$ is the radius of the cluster, and (ii) that cluster size scales as $P^2$ \cite{hagena}, the scattered light signal should scale as $P^3$. The dependence of the scattered light signal with pressure that we measured was found to be consistent with this scaling \cite{xray}, thus confirming that Hagena's law is applicable to our cluster source.   
   
(H$_2$O)$_n$ and (D$_2$O)$_n$ clusters were produced in our experiments by bubbling either He or Ar gas through liquid water at the stagnation chamber.  To estimate the size of the water clusters, we resorted to the correspondence principle formulated by Hagena in estimating the sizes of metal clusters produced in supersonic jet expansion \cite{hagena-ZPD}. The modified Hagena parameter is 
\begin{equation}
\Gamma^*=\Gamma ~r_{ch}^{2.15} ~T_{ch}^{1.29},
\end{equation}
where 
\begin{eqnarray}
r_{ch}=(m/\rho)^{1/3},\\
T_{ch}=\Delta h_s^o /k. 
\end{eqnarray}
In the above, $m$ represents the mass, $\rho$ the solid density, and $\Delta$ h$_s^o$ the enthalpy of sublimation of the cluster system. We use the correspondence principle in conjunction with the argon cluster size measurements using Rayleigh scattering \cite{xray} in order to obtain an estimate of the size of 
water clusters which seed high pressure helium gas under similar experimental conditions.
Using these equations we derive the ratio of the Hagena
parameter for Ar and for H$_2$O (D$_2$O). For water, $\Delta$ h$_s^o$= 50.9 kJ/mol, $\rho$=917 kg m$^{-3}$ and for Ar,  $\Delta$ h$_s^o$= 7.7 kJ/mol, $\rho$=1707 kg m$^{-3}$, yielding 
\begin{equation}
{{\Gamma^* (Ar)}\over {\Gamma^* (H_2O)}}= 0.08 \cdot  {{P(Ar)}\over{P (H_2O)}},
\end{equation}
and mean cluster sizes are given by
\begin{equation}
<N(H_2O)>= 154  {\Big [}{{P (H_2O)}\over{P(Ar)}}{\Big ]}^2 \cdot <N(Ar)>.
\end{equation}
The vapor pressure of water is 0.03 atmosphere at 300 K temperature.
Taking the partial pressure for water in the supersonic beam that is seeded with 3 atmosphere of He to be
0.03 atmosphere, and from our earlier results that the mean size of Ar cluster is $\sim$4000 atoms 
at 3 atmosphere pressure, we estimate the corresponding size for the water clusters to be about 60 atoms.

We note that recourse to the correspondence principle has also been made in earlier studies of large (HI)$_n$ clusters \cite{springate} as well as carbon-containing clusters (CO$_2$)$_n$ and (C$_3$H$_8$)$_n$ \cite{muller}. Differences in the nature of bonding in these cluster species, as compared to the bonding in rare gas clusters, might be considered to be of some importance in determining the degree of clustering that may be attained, and in determining the applicability of the correspondence principle to clustering that relies on hydrogen bonding. Direct measurements of water cluster sizes have been made by mass spectrometry in apparatus with a supersonic nozzle of the type we use \cite{ahmed}. Variation of the average cluster size with stagnation pressure has enabled a scaling law to be established \cite{Dubov}. Comparison of the cluster sizes we estimate shows good accord with the results of direct measurements.

Velocity analysis and detection of charged particles produced in the laser-cluster interaction were carried out by means of a 58-cm long time-of-flight spectrometer and a channeltron/microchannel plate, respectively. Charge state analysis of velocity-discriminated charged particles was accomplished using a retarding potential energy analyzer (RPA) comprising three electrodes, each with 90\% transmission nickel mesh, placed in front of the detector. Grounded plates of the same type were placed at a separation of 5 mm on both sides of the central retarding plate so that the flight-tube remained field-free. The retarding plate could be held at a potential of up to $\pm$5 kV. The stable high-voltage supply was remote-controlled by means of a 0-9 V input signal that was generated using a 12-bit digital-to-analog converter (DAC). The DAC was used to ramp the voltage on the RPA such a two-dimensional matrix of TOF spectra could be constructed from a series of spectra taken at different values of voltage on the RPA. The column number of such a matrix represents the time-of-arrival of the ion, and the row number corresponds to the voltage applied to the RPA. The ion intensity distribution in time was transformed to a distribution in energy, and adjacent-averaged to reduce noise. The 2D distribution thus obtained was then differentiated with respect to RPA voltage to obtain an energy spectrum. The process of numerical differentiation was, in this case, equivalent to taking the difference between adjacent voltage spectra, and this difference corresponds to the yield of ions at energies covered by this voltage interval. 

The 2D spectrum thus represents charge-resolved ion yields as a function of kinetic energy, with the novelty of our approach being, as shown in the following, that a single spectrum displays the energy distribution for the plethora of charge states that are produced in the course of the laser-cluster interaction. The technique is easily applicable to hetero-nuclear clusters. In the experiments presented here we show that one can distinguish the protons very clearly from the O$^{q+}$ produced in the water clusters. Use of pulse height analysis techniques on the other hand for distinguishing atomic ions in this case on the other hand would have not succeeded. 

The laser pulses in these experiments were produced by a 100 femtosecond, 55 mJ Titanium:sapphire laser. The laser was focused into the vacuum chamber with a 25 cm focal length plano-convex lens to produce intensities up to 10$^{16}$ W cm$^{-2}$. The experiments reported in this study were performed at intensities ranging from 10$^{14}$ to 10$^{16}$ W cm$^{-2}$.
 
\section{Results and discussion}

\subsection{Argon clusters}

A typical raw TOF spectrum for $Ar_{40000}$ is shown in Figure 2. This spectrum was acquired at a laser intensity of 8$\times$10$^{15}$ W cm$^{-2}$, with the laser polarization vector aligned along the axis of the TOF spectrometer. The stagnation pressure was 10 bar. The narrow fast peak that is observed at very short ($<$0.2$\mu$s) flight times is due to electrons/photons. The remaining structure, beyond flight times of 0.5$\mu$s, is ascribed to energetic ions. It was confirmed that both peaks are present only when there is spatial and temporal overlap between the cluster beam and the laser beam. The corresponding ion energy spectrum, $f(E)$, that was obtained from the time-of-flight spectrum, $f(t)$, using the relation 
\begin{equation}
f(E) = f(t) [dE/dt]^{-1}, 
\end{equation}
is shown in Figure 3. The general features of this spectrum are in good accord with a spectrum reported earlier \cite{ditmire3}. The highest energy seen in this spectrum is about 500 keV, and the mean energy of the ions,  
 \begin{equation}
\bar E = \frac{\int{E f(e) dE}}{f(E) dE},
\end{equation}
is about 20 keV. The highest ion energy is determined from the minimum in the raw TOF spectrum shown in Figure 2, between the ion and the electron/photon features. Note that the minimum in the TOF spectrum is not zero, implying that there is an overlap between the two peaks. The determination of the maximum ion energy is, therefore, a conservative estimate - the highest energy ions probably have energies higher than 500 keV. The digital oscilloscope used in these experiments also had a finite record-length (5000 points in this case) and this determined the low-energy cut-off in the spectrum. For our experimental setup, this cut-off for argon ions is at 180 eV. The averaging over ion energies therefore does not include ions with energies lower than this limit. As we will show in the following, fragmentation of $Ar_n$ clusters gives rise to argon ion charge states up to 8+ and each of the ejected electrons typically possesses an energy of the order of 0.5 keV under conditions prevailing when data in Figure 2 was acquired. Taken together with the potential energy associated with the highly-charged argon ions, it is clear that in the interaction of the laser with $Ar_{40000}$ clusters at an intensity of 8$\times$10$^{15}$ W cm$^{-2}$, the total energy that is deposited is of the order of 25 keV per argon atom. In the case of larger clusters, such as $Xe_{150000}$, irradiated by similar laser fields, energy deposition as high as 100 keV per atom has been measured \cite{Xe}.  Figure 3 also shows the energy profile obtained in the case of a somewhat smaller cluster ($Ar_{2000}$) obtained by using a stagnation pressure of 2 bar. Values of the peak and average energies obtained in this case are both significantly smaller than those obtained in the case of $Ar_{40000}$, although the overall morphology of the ion energy distribution function remains very similar. The asymmetry in ion emission that was recently reported in the case of $Ar_{40000}$ \cite{prl} is also observed when $Ar_{2000}$ clusters undergo explosion.
 
Energy spectra were also acquired with various positive voltages applied on the RPA. Results for $Ar_{40000}$ obtained with 200 V and 400 V are shown in Figure 4. When a voltage $V$ is applied to the retarding plate, all ions with charge q and energy less than $qV$ are unable to overcome the potential barrier, and cannot reach the detector. The TOF of ions with energies greater than the cut-off remains unaffected because the ions spend only a very small fraction of their total flight time in the RPA assembly. Therefore, the spectrum that is obtained shows a series of steps corresponding to values of ionic charge state $q$. The height of each step indicates the yield of the charge state $q$ at energy $qV$. The spectra shown in Figure 4 show steps for $Ar^+$ to $Ar^{7+}$ when 200 V is applied to the RPA while $Ar^{4+}$ to $Ar^{8+}$ can be distinguished when the RPA voltage is 400 V. 

In order to reinforce the utility of our two-dimensional technique, we reiterate that conventional, one-dimensional TOF spectrometry would not be able to yield the type of information that is discussed above. If we were to apply normal Wiley-McLaren conditions to our TOF spectrometer, using extraction voltages of $\pm$250 V, $Ar^+$ ions possessing either thermal energy, or 10 keV energy, would have mean flight times of 12.6 $\mu$s or 2.6 $\mu$s, respectively. For $Ar^{2+}$ ions, the corresponding mean flight times would be 8.9 $\mu$s (thermal energy ions) and 2.6 $\mu$s (10 keV ions). In the case of $Ar^{10+}$ ions, the corresponding mean flight times would be 3.9 $\mu$s and 2.2 $\mu$s. Even if we disregard the inevitable temporal widths of the TOF peaks associated with each of these ion species, it is clear that conventional TOF spectrometry would not enable us to disentangle the plethora of ions that would appear in the flight time region of 2.5 $\mu$s. 

Figure 5 shows a two-dimensional spectrum of $Ar_{40000}$ constructed from a series of spectra taken with different RPA voltages.We have reported this spectrum earlier \cite{prl} in connection with the discovery of asymmetric ion emission upon fragmentation of $Ar_{40000}$ clusters.  As noted above, the 2D spectrum represents charge-resolved ion yields as a function of ion kinetic energy. As the $y$-axis also represents the energy per charge state on the ion, the various charge states appear in our 2D spectrum as lines whose slope is $q^{-1}$. In such a depiction, a linear dependence of ion energy on charge state, as prevailing wisdom would expect if the hydrodynamic model were to provide a realistic representation of the laser-$Ar_{40000}$ dynamics, would cause the mean energies to lie on a horizontal line. Our 2D-TOF data shows that this is clearly not the case. In fact, analysis shows that the observed dependence is approximately quadratic, indicating a Coulomb explosion according to prevailing wisdom. This finding is consistent with the results of ion energy measurements conducted by Lezius {\it et al.} \cite{normand} on somewhat larger argon clusters where the authors had concluded that the clusters undergo Coulomb explosion. In both studies, multiply charged ions are observed up to $Ar^{8+}$; beyond that the spectrum could not be resolved into constituent charge states. We attribute this to the significant difference in the ionization energies of $Ar^{7+}$ (143 eV) and $Ar^{8+}$ (422 eV). While higher charge states are certainly produced, as evidenced by the appearance of recombination lines in the emission spectrum, the fraction of ions in these highly charged states is likely to be very small.     

\subsection{Water clusters}

We have conducted experiments on molecular clusters, $(H_2O)_n$ and $(D_2O)_n$, that were produced by passing high-pressure helium or argon gas through a reservoir of high-purity $H_2O$ or $D_2O$. As noted above, we estimate that the degree of clustering that we obtain in our experiments is such that $n\sim$60. The formation of these water clusters is dependent upon hydrogen bonding and polarizable intramolecular forces. In aggregates of water molecules, the electric field from nearest neighbors induces alterations in electronic configuration that, in turn, further affect interactions with other water molecules. This form of cooperativity has been studied by Xu {\it et al.} \cite{xu} who describe the process when a water molecule forms a hydrogen bond and undergoes an internal rearrangement of charge density such that it leads to an overall strengthening of other hydrogen bonds that are formed by the same water molecule. This mutual enhancement of hydrogen bond strength accounts for the observed propensity of water to aggregate into clusters. 

Typical ion energy spectra that we obtained in our experiments by applying two different RPA voltages on the spectrometer are depicted in Figure 6. Several charge states of oxygen ions are readily identified in both spectra: in the case of 500 V on the RPA, note the sharp onset of peaks at 1000V, 1500V, and so on, that indicate $O^{2+}$, $O^{3+}$ ions, and so on. Charge states as high as 6+ are observed (helium-like oxygen). As in the case of argon ions, the contribution from ionization states resulting from K-shell vacancies in oxygen are probably too small to be of much import in the overall dynamics. We note that peaks due to $H^+$ ions ostensibly appear at the highest end of the spectra that are shown; this is, obviously, a manifestation of both the low mass (short-duration TOF) of protons as well as the energies that they possess. Of course, the disassembly of highly-charged water clusters will give rise to energetic ions, and largest kinetic energies will be carried away by the lowest-mass protons. 

Figure 7 shows a 2D spectrum that is obtained for $(H_2O)_n$ clusters. It is clear from the 
spectrum that our 2D-TOF technique shows a clear advantage in distinguishing atomic ions of different mass. The energy scale in the spectrum is calculated from the arrival time taking the proton mass into account. So to actually obtain the oxygen ion energies one has to multiply the energy scale by 16, the mass of oxygen atom. The spectrum clearly shows evidence for $O^{q+}$ ions, with $q$ as large as 6. Careful analysis reveals, most unexpectedly, that the energy dependence of different values of $q$ for $O^{q+}$ ions actually scales linearly with $q$. This is most clearly seen by considering, for example, the maximum values of ion energy on each of the straight lines in Fig. 7 that represent 
different charge states of oxygen. It is clear from the spectrum that the maximum ion energy values 
fall on a line that is parallel to the x-axis, indicating that $O^{q+}$ ion energies scales linearly with $q$.  This is surprising if we recall the earlier discussion on the two prevalent models that seek to describe laser-cluster explosion dynamics: hydrodynamic expansion and Coulomb explosion. Conventional wisdom based on these two models demands that relatively large clusters would follow, at least qualitatively, the predictions of the uniform density hydrodynamic expansion model whereas smaller clusters would tend to behave according to the expectations of the Coulomb explosion model. As noted in the Introduction, in the former case, disassembly of clusters would yield highly charged ions whose energy scales as $q$. In the latter case, the corresponding energy dependence would be close to $q^2$. Our experiments on $Ar_n$ clusters, with $n$ as large as 40000, reveal an ion energy dependence that scales approximately as $q^2$. On the other hand, experiments that we conducted on $Xe_n$ clusters \cite{prl}, with $n\sim$ 150000, yielded definite evidence that the explosion dynamics in this case most definitely occurred in the hydrodynamic regime (even though our results indicated the need to introduce a two-dimensional component into the hydrodynamic expansion model in order to rationalize the asymmetry in the ion emission process that was discovered.)  The present results indicate a linear dependence of ion energy of charge state in the case of relatively small water clusters, (H$_2$O)$_n$ and (D$_2$O)$_n$, $n\sim$60, that have been produced by seeing high pressure He and Ar gases. From the estimates of the size of water clusters presented earlier, the water clusters in our experiments are very much smaller in size than the rare gas clusters that are produced at the stagnation pressure utilized in our experiments. The estimation that the clusters are smaller in size is also augmented by the fact that the maximum ion energies observed in water clusters are lower by nearly two orders in magnitude. If the cluster size was bigger, the kinetic energy release would have been much larger. These finding raise the question whether the linear energy-charge relationship that is measured for water clusters does, in fact, indicate  hydrodynamic expansion.  

\subsection{Coulomb explosion versus hydrodynamic expansion}

An important consequence of the prevailing wisdom regarding cluster explosion dynamics, as already discussed in outline in the Introduction, dictates that relatively large clusters follow the predictions of the hydrodynamic expansion model whereas smaller clusters behave according to the expectations of the Coulomb explosion model. In the former case, disassembly of clusters whose diameter $a>>\zeta_e,\delta_e$, would yield highly charged ions whose energy scales as $q$, the ionic charge. Specifically, the mean energy, $\bar E$, and the mean charge of fragment ions, $\bar q$, would be related by
\begin{equation}
\bar E = \frac{3}{2}\bar q k_B T_e \propto \bar q,
\end{equation}
where $k_B$ is the Boltzmann constant and $T_e$ is the electron temperature which, it is assumed, depends only weakly on $\bar q$. 

On the other hand, the Coulomb explosion scenario demands that the corresponding energy dependence be $q^2$. In practice, a cluster comprising $N$ atoms would be expected to explode into ions that possess a range of values of charge and energy that may be represented by $q_j$ and $E_j$, respectively, for the $j^{th}$ ion. The mean energy and mean charge state would then be specifically related by
\begin{equation}
\bar E = \frac{\bar {q^2}}{N} \sum_{i=1}^{N=1} \sum_{j=i+1}^N \frac{1}{\left |R_i - R_j\right |} \propto \bar q^2,
\end{equation}
where $R_i$ denotes the initial position of the $i^{th}$ ion. 

The dependence of the ion energy on $q$ has constituted the criterion that have been used in previous studies to experimentally disentangle cluster explosion dynamics in the two regimes \cite{normand}, despite the fact that experiments give direct access only to $\bar E$ and not to $\bar q$. Experiments provide access to information on specific charge states $q$ and associated energies $E$ of individual ions that are formed upon explosion of the parent cluster. Ishikawa and Blenski \cite{blenski} have carried out Monte Carlo classical particle-dynamics simulations of small clusters of Ar and Xe (up to 147 atoms) and have determined values of $q$ and $E$ for specific subshells in the clusters. Their results show that Coulomb explosion of relatively small clusters like $Xe_{147}$ give rise to a distinctly {\em linear} dependence of ion energy on high charge states which, in prevailing wisdom, would be attributed to hydrodynamic expansion. In other words, the results of the simulations question the wisdom of using the simple experimental recipe of ion energy being proportional to either $q$ or to $q^2$ in order to disentangle Coulomb explosion from hydrodynamic expansion. 

Relatively little work has been reported on the interaction of intense laser light with molecular clusters \cite{springate,muller,castleman} and it has relied exclusively on the application of one-dimensional TOF methodology to clusters of modest size. The application of 2D-TOF methodology to atomic \cite{ditmire1,prl} and molecular clusters opens new vistas for such research; as the present results indicate, it also questions prevailing wisdom regarding cluster explosion dynamics. The method certainly establishes the fact that water clusters formed by seeding with high pressure argon gas can be used to produce energetic and highly-charged ions of oxygen. Moreover, our results appear to offer experimental confirmation of what the simulation studies of Ishikawa and Blenski \cite{blenski} predicted (but have not, hitherto, been taken proper cognizance of): charge state dependences of ion energy do not constitute an unambiguous enough experimental signature of cluster explosion regime.

\begin{figure}
\caption{Schematic representation of the experimental apparatus for the measurement of ion energy spectra upon disassembly of rare gas and rare gas + water clusters. HV: High voltage supply for the retarding potential analyzer, DAC: Digital-to-Analog-Convertor, DSO: Digital Storage Oscilloscope, GPIB: General Purpose Interface Board.}
\end{figure}

\begin{figure}
\caption{Typical raw output signal from the microchannel plate (MCP) detector near the laser arrival time (t=0) obtained when $Ar_{40000}$ clusters are irradiated by a laser intensity of 8$\times$10$^{15}$ W cm$^{-2}$. The feature occuring at very short times ($<$0.2 $\mu$s) is ascribed to photons and electrons. The features beyond 0.5 $\mu$s are due to energetic ions.}
\end{figure}

\begin{figure}
\caption{Ion energy distributions of $Ar_{40000}$ and $Ar_{2000}$ clusters irradiated by a laser intensity of 8$\times$10$^{15}$ W cm$^{-2}$, with the laser polarization vector directed parallel to the axis of the time-of-flight spectrometer.}
\end{figure}

\begin{figure}
\caption{Ion energy distribution of $Ar_{40000}$ clusters irradiated by a laser intensity of 8$\times$10$^{15}$ W cm$^{-2}$ with two different voltages applied on the retarding potential analyser (RPA). The spectra are mutually displaced along the vertical scale for clarity.}  
\end{figure}

\begin{figure}
\caption{2D-TOF spectrum showing charged-resolved ion energy distributions of $Ar_{40000}$ clusters. Each charge state $q$ is depicted as a line with slope $q^{-1}$. The ion energy depends on ion charge state as $\sim q^2$.}
\end{figure}

\begin{figure}
\caption{Energy distribution of ions from water clusters irradiated by a laser intensity of 8$\times$10$^{15}$ W cm$^{-2}$ with voltages of 500 V and 600 V applied on the retarding potential analyser (RPA). The spectra are mutually displaced along the vertical scale for clarity.  Note that the time-to-energy conversion of the raw TOF spectra is with respect to oxygen ions in this depiction.}
\end{figure}

\begin{figure}
\caption{2D-TOF spectrum of water clusters showing charged-resolved ion energy distributions. Note that the time-to-energy conversion of the raw TOF spectra is with respect to $H^+$ ions in this depiction. The energies of $O^{q+}$ ions depend almost linearly on ion charge state $q$.}
\end{figure}

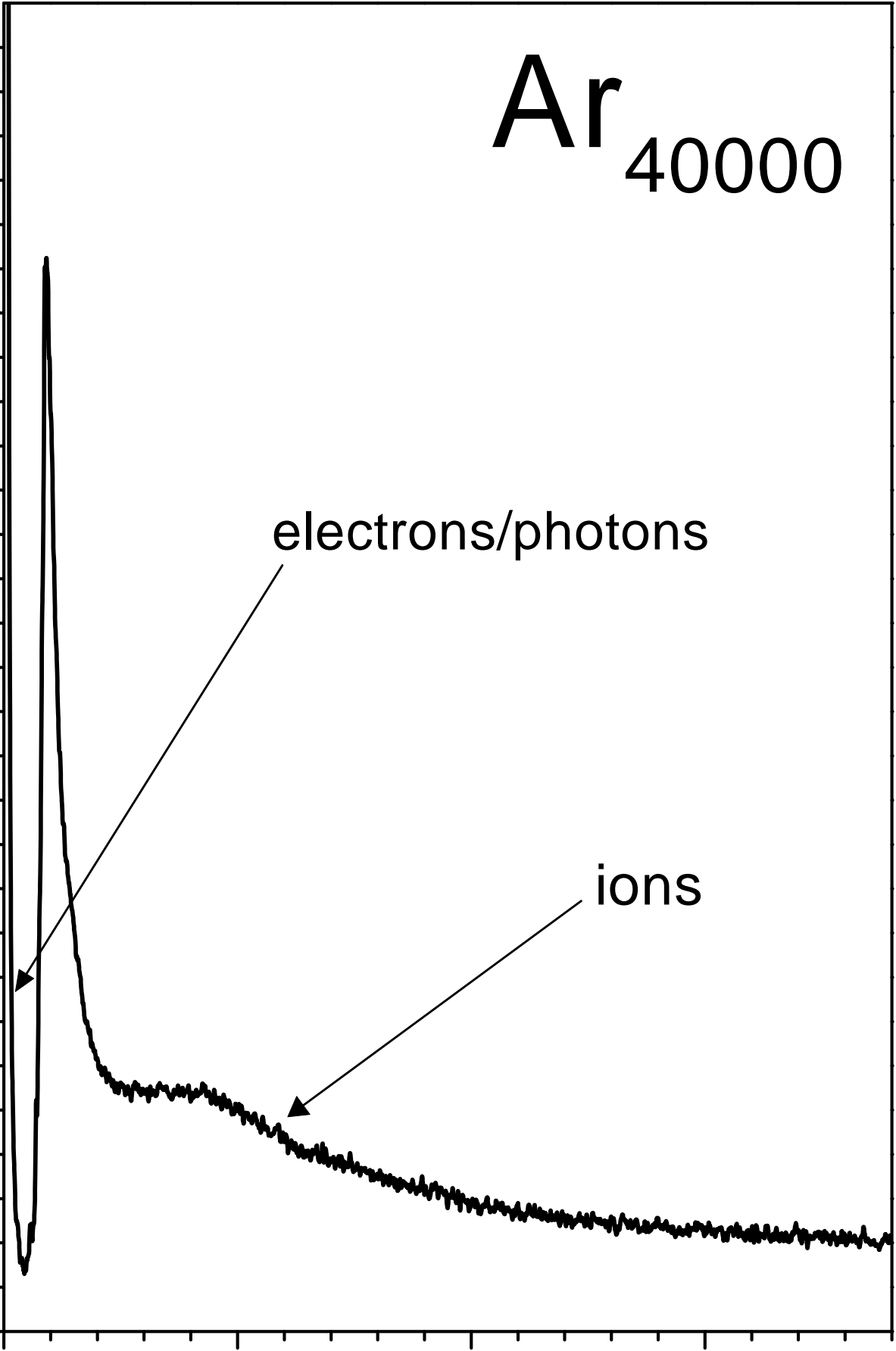
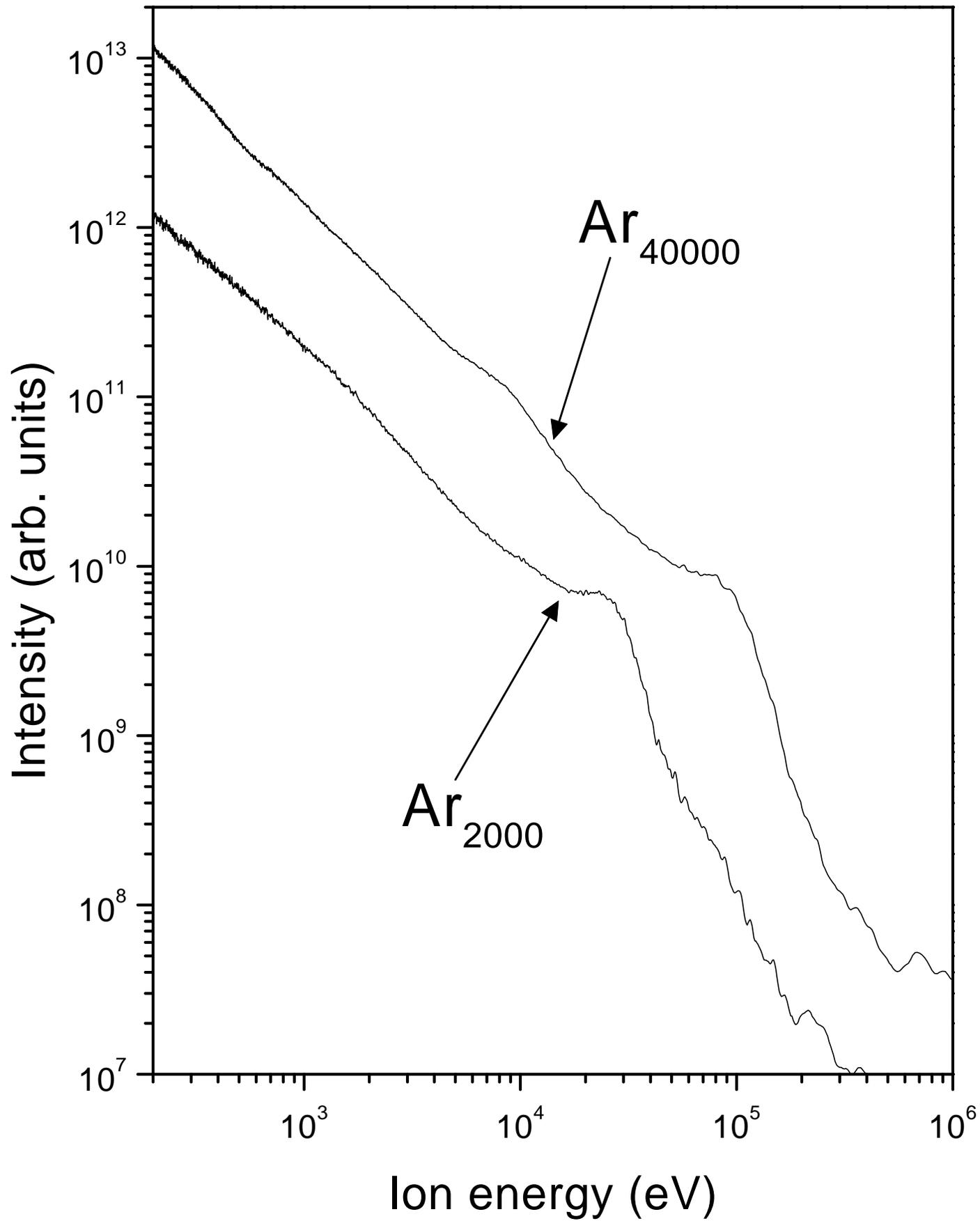
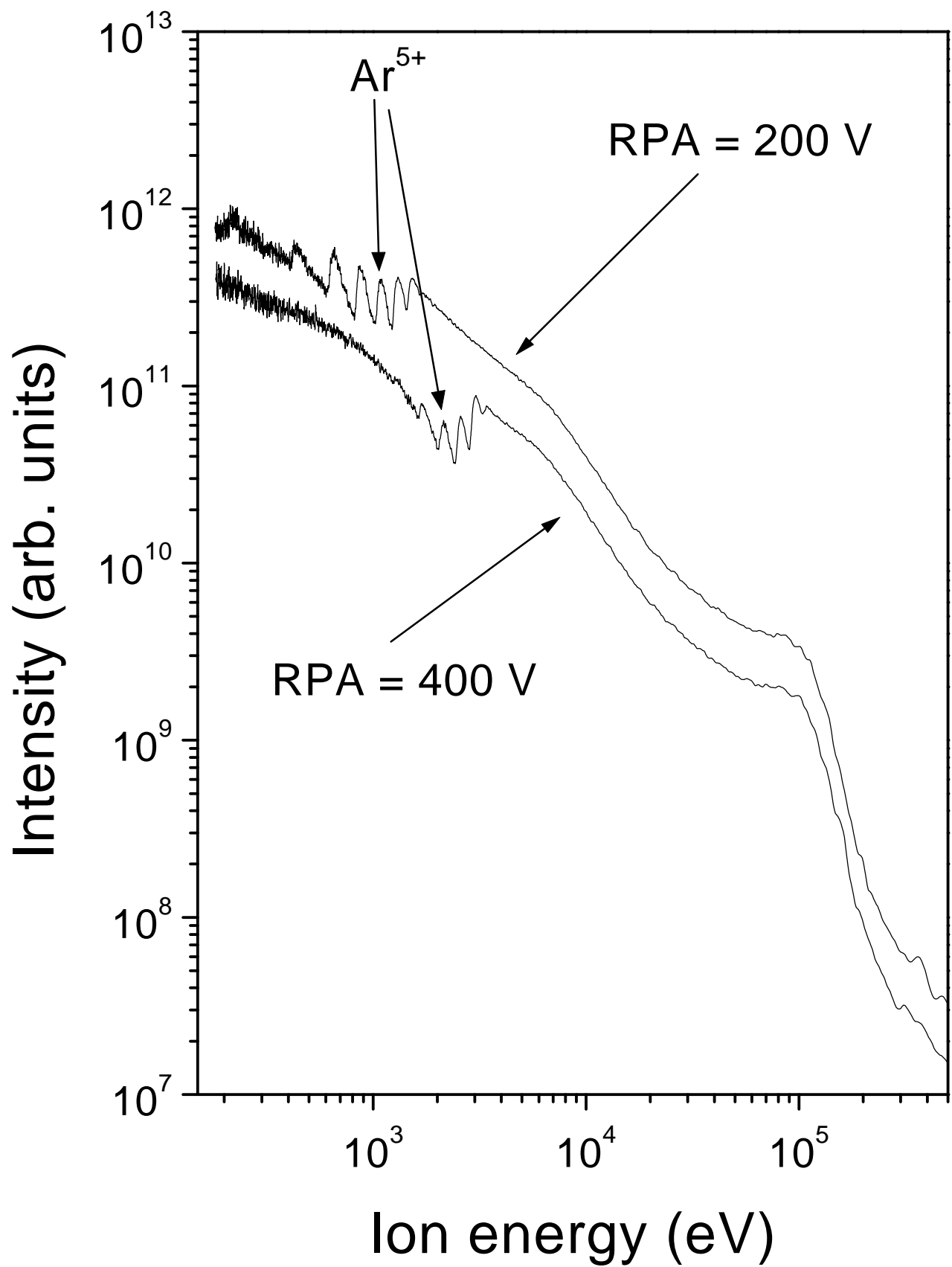
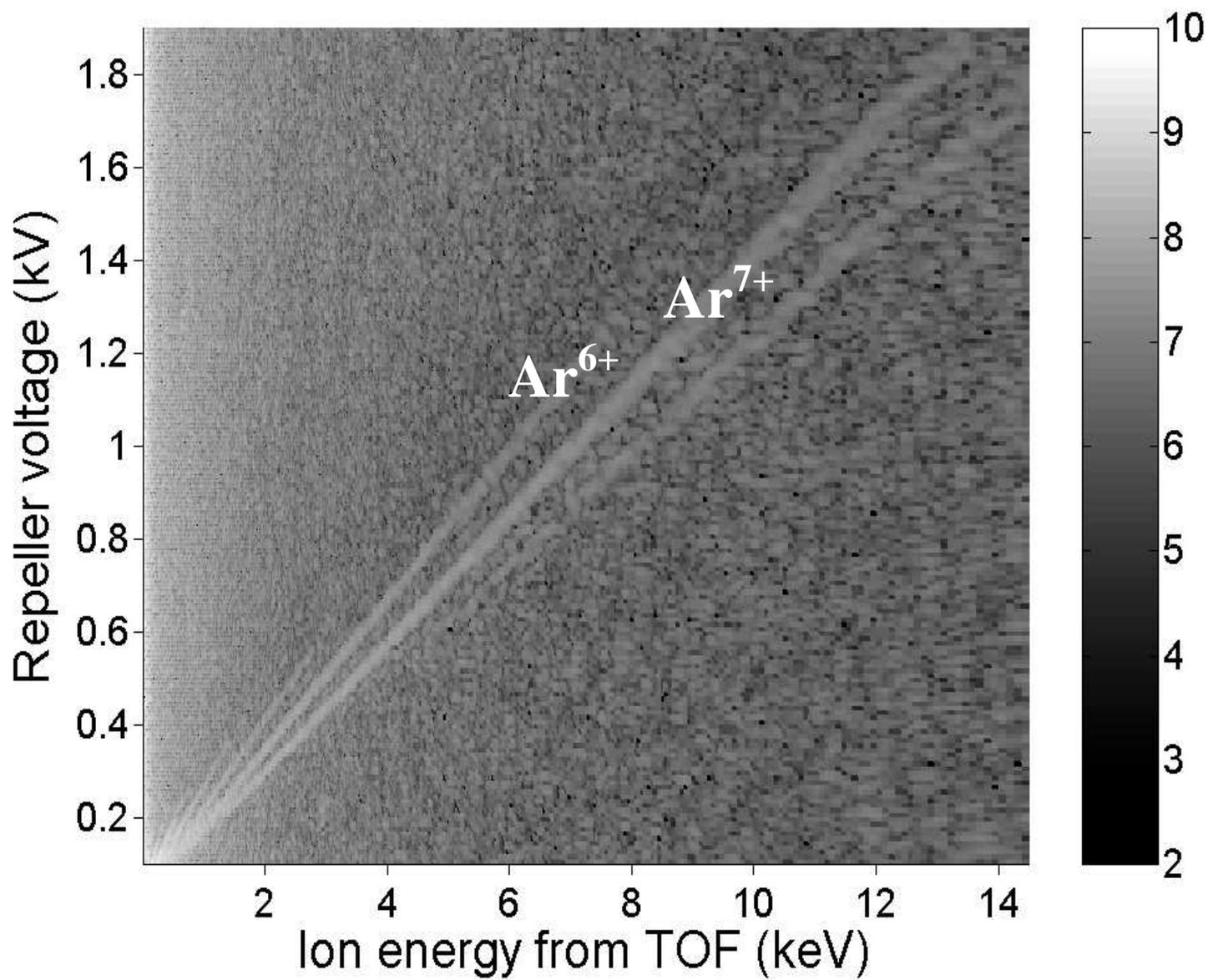
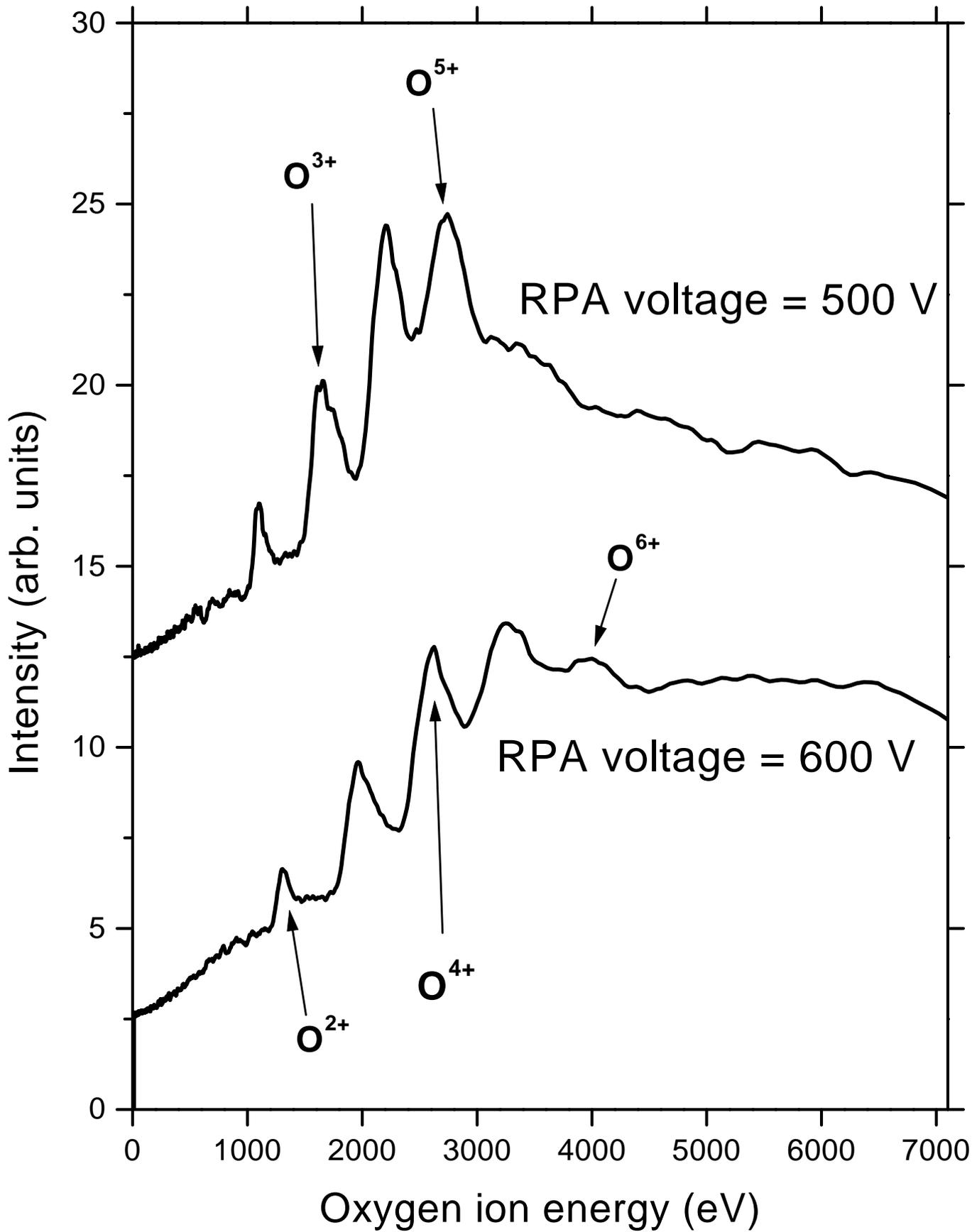
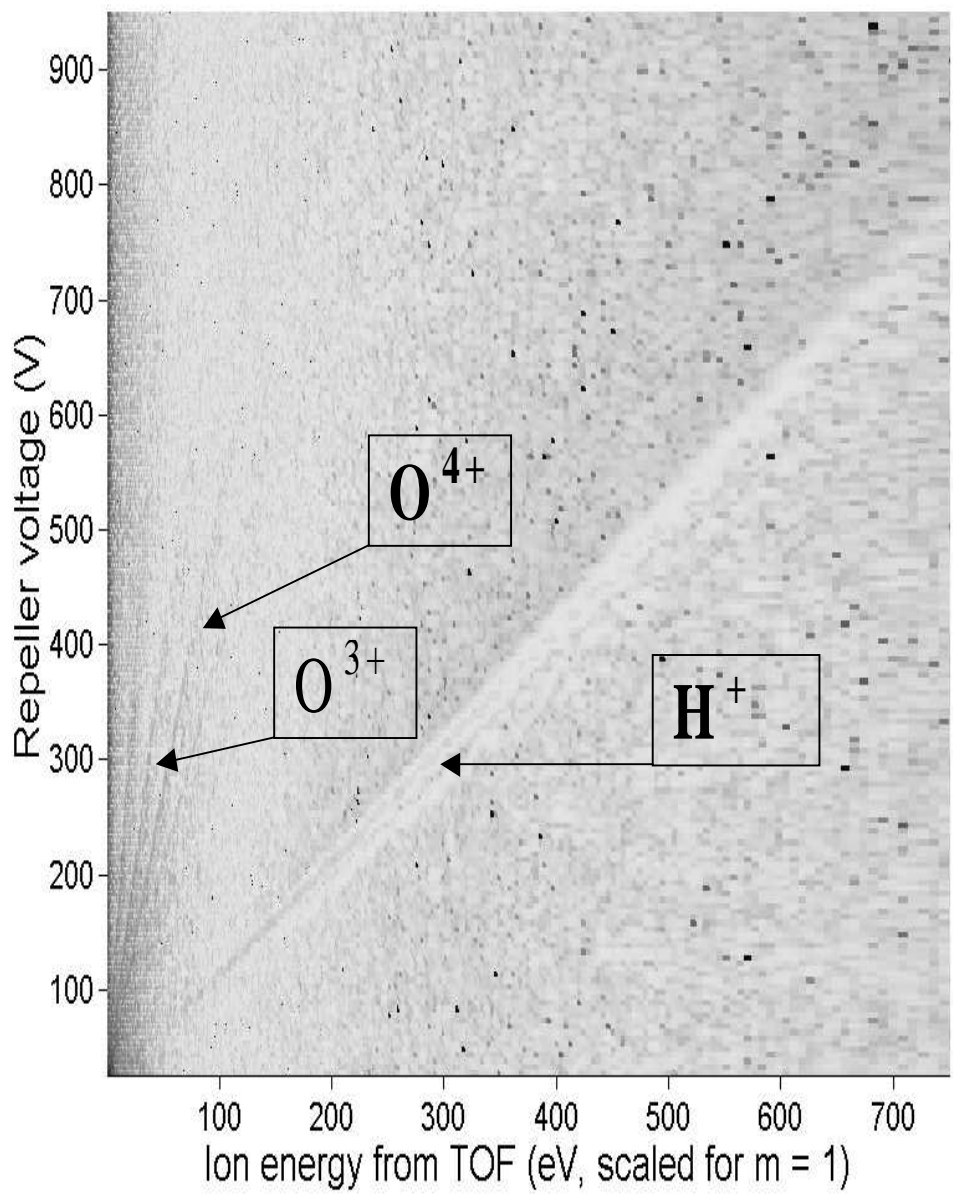
\end{document}